\documentclass[twocolumn, showpacs, longbibliography, prl,aps, amsmath, amssymb,superscriptaddress]{revtex4-2}
\usepackage{amsmath}
\usepackage{amsfonts}
\usepackage{graphicx}
\usepackage[colorlinks,linkcolor=blue,citecolor=blue,urlcolor=blue]{hyperref}
\usepackage{float}
\usepackage[normalem]{ulem}
\usepackage[switch,columnwise]{lineno}
\usepackage[dvipsnames]{xcolor}
\definecolor{Green2}{rgb}{0, 0.4, 0}

\begin{document}
\title{Steady-state superfluidity of light in a tunable cavity at room temperature}

\author{G. Keijsers}
\affiliation {Center for Nanophotonics, AMOLF, Science Park 104, 1098 XG Amsterdam, The Netherlands}

\author{Z. Geng}
\affiliation {Center for Nanophotonics, AMOLF, Science Park 104, 1098 XG Amsterdam, The Netherlands}

\author{K. J. H. Peters}
\affiliation {Center for Nanophotonics, AMOLF, Science Park 104, 1098 XG Amsterdam, The Netherlands}

\author{M. Wouters}
\affiliation {TQC, Universiteit Antwerpen, Universiteitsplein 1, B-2610 Antwerpen, Belgium}

\author{S. R. K. Rodriguez}  \email{s.rodriguez@amolf.nl}
\affiliation {Center for Nanophotonics, AMOLF, Science Park 104, 1098 XG Amsterdam, The Netherlands}

\begin{abstract}
\noindent Light in a nonlinear cavity is expected to flow without friction --- like a superfluid --- under certain conditions.  Until now, part-light part-matter (i.e., polariton) superfluids have been observed either at liquid helium temperatures in steady state, or at room temperature for sub-picosecond timescales. Here we report signatures of superfluid cavity photons (not polaritons) for the first time. When launching a photon fluid against a defect, we observe a suppression of backscattering above a critical intensity and below a critical velocity. Room-temperature and steady-state photon superfluidity emerges thanks to the strong thermo-optical nonlinearity of our oil-filled cavity. Numerical simulations qualitatively reproduce our experimental observations, and reveal how a viscous photon fluid reorganizes into a superfluid within the thermal relaxation time of the oil. Our results establish thermo-optical nonlinear cavities as platforms for probing photon superfluidity at room temperature, and offer perspectives for exploring superfluidity in arbitrary potential landscapes using structured mirrors.
\end{abstract}

\date{\today}
\maketitle

\noindent\textbf{INTRODUCTION} \\
Superfluidity --- frictionless flow --- is a striking manifestation of macroscopic quantum coherence. Superfluidity was discovered a little over 80 years ago, when helium was cooled below 2.2 K~\cite{Kapitza38, Misener1938}. Since then, there has been great interest in understanding the fascinating properties of superfluids, and inducing superfluidity at increasingly higher temperatures~\cite{pitaevskii2016bose}. Experiments with liquid helium~\cite{Rayfield66, Phillips74, Castelijns86} and atomic Bose-Einstein condensates~\cite{Raman1999, Onofrio00, Dalibard12} have revealed the rich phenomenology associated with superfluidity. However, room-temperature superfluidity in those systems remains a distant goal.

About 20 years ago, a new frontier of physics emerged from the following recognition: the Lugiato-Lefever equation for light in a nonlinear cavity resembles the Gross-Pitaevskii equation for the order parameter of a superfluid~\cite{Lugiato87, Chiao99, Bolda01, Carusotto04}. A connection between optical systems and superfluids was further established by the realization that, above a critical light intensity and below a critical photon velocity, light can flow  into an obstacle without scattering.  Crucially, the cavity needs to contain a material with intensity-dependent refractive index for this effect to emerge. Such a  material mediates effective photon-photon interactions, which are essential for superfluidity according to Landau’s criterion~\cite{Landau1941}. Until now,  photon superfluidity as conceived in Refs.~\onlinecite{Chiao99, Bolda01} has never been realized. However, superfluid behavior has been observed in two optical systems: i) polaritons in semiconductor cavities~\cite{Amo2009, Lagoudakis2008, Sanvitto2010, Nardin11, Amo11,  CarusottoRMP, Ballarini20}, and ii) optical beams propagating in nonlinear media~\cite{Wan2007, Faccio15, Vocke16, Michel18, Glorieux18}.

Superfluidity of cavity polaritons has been observed at liquid helium temperatures in steady-state~\cite{Amo2009, Lagoudakis2008, Sanvitto2010, Nardin11, CarusottoRMP}, and recently at room-temperature but for sub-picosecond timescales only~\cite{Lerario2017}.  Room-temperature steady-state polariton superfluidity has never been realized. This is likely due to an inherent trade-off between  nonlinearity strength $U$  and operation temperature in polariton systems. Both of these quantities are directly related to the exciton binding energy $E_b$, albeit in inverse ways. While $U \propto E_b^{-1}$, the existence of polaritons hinges on $E_b > k_B T$ with $k_B T$ the energy of thermal fluctuations. Hence, room-temperature polaritons based on large $E_b$ excitons generally result in weak nonlinearities. The other configuration used to explore fluids of light so far, where an optical beam propagates through a nonlinear medium without any cavity involved, has allowed for the observation of superfluid-like behavior at room temperature~\cite{Wan2007, Faccio15, Vocke16, Michel18, Glorieux18, Carusotto14}. However, it should be stressed that the interpretation of superfluidity in those experiments hinges on mapping  the spatial coordinate along the beam propagation direction to the temporal dimension in the Gross-Pitaevskii equation. Thus, while interesting in their own right, those analog systems miss the important temporal degree of freedom characteristic of all fluids propagating in space and time.

A cavity with thermo-optical nonlinearity was recently proposed as an alternative system for realizing the long-sought photon superfluid~\cite{Alaeian17}. Thermo-optical nonlinearities are routinely observed at room temperature and under continuous driving~\cite{Lipson04, Carmon04, Notomi05, Priem05,  Geng_PRL2020}, which is encouraging for experiments. However, the non-instantaneous and non-local character of thermo-optical nonlinearities are expected to reduce the critical velocity for superfluidity, and to limit its observation to an extremely narrow frequency range~\cite{Alaeian17}. In particular, for moderate light intensities the critical velocity is limited by the energy of a roton-like minimum appearing in the dispersion of excitations in non-local media~\cite{Alaeian17}.  Above a critical intensity, however, the excitation spectrum becomes gapped and superfluidity can emerge. This is precisely the regime we explore below.

Here we report signatures of superfluid cavity photons in steady state and at room temperature. We investigate a laser-driven tunable cavity filled with olive oil, which has a strong thermo-optical nonlinearity. By varying the intensity and momentum distribution of the incident laser, as well as the cavity length, we control the density and velocity of the photon fluid in the cavity. The transition to superfluidity is studied by measuring the transmitted intensity and momentum distribution for different flow conditions, when launching a photon fluid against a  naturally-occurring defect in one of the cavity mirrors. Superfluidity is evidenced by the disappearance of scattering from the defect  above a critical incident power and below a critical velocity.  Our interpretation in terms of photon superfluidity is confirmed by numerical simulations  based on a driven-dissipative Gross-Pitaevskii equation for a photon fluid coupled to a temperature field. The simulations also show that, while the thermal character of the nonlinearity influences the time scale in which superfluidity emerges, the steady-state density profile at large intensities is similar to the one observed for a cavity with instantaneous Kerr nonlinearity.\\

\begin{figure}[!t]
	\includegraphics[width=\columnwidth]{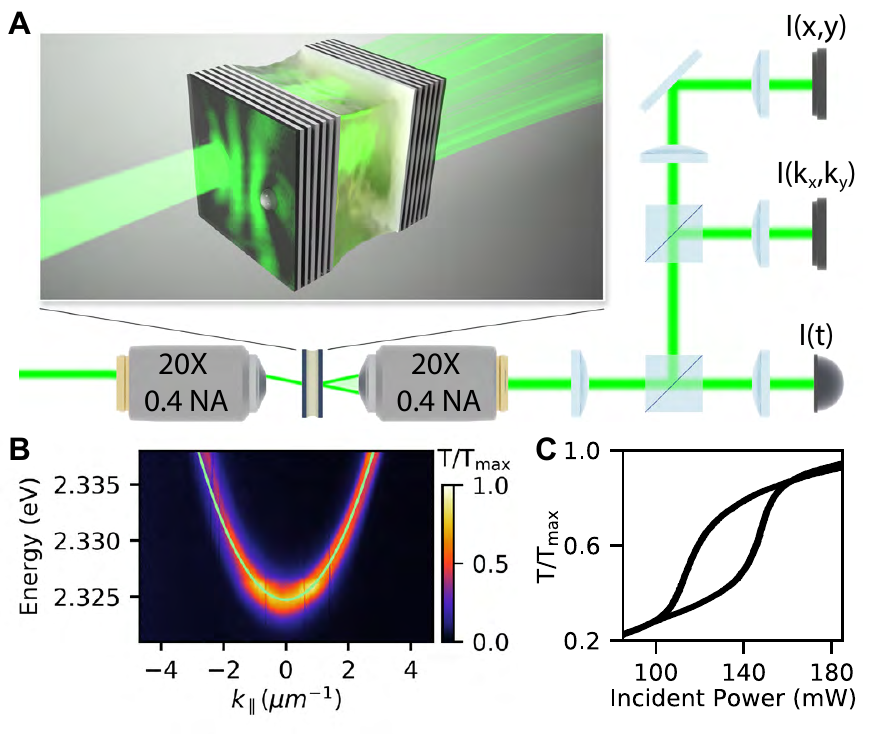}
	\caption{\label{fig:1}\textbf{Experimental set-up, photonic dispersion, and optical bistability.} (\textbf{A}) Illustration of our experiment, involving an oil-filled optical cavity driven by a continuous wave laser. The cavity length is controlled with a piezoelectric actuator holding one of the mirrors. The transmitted power is time-resolved with a photodetector, or measured in position and momentum space with cameras. (\textbf{B}) White light transmission measurements of our oil-filled cavity. The measured dispersion relation is fitted by a parabola, indicating the cavity photons behave as free particles with an effective mass $m^{\ast}= 2.064 \times 10^{-35}$ kg. (\textbf{C}) Transmitted power measured at $k_\parallel=0$ when ramping the laser power up and down, evidencing optical bistability due to effective photon-photon interactions.}
\end{figure}

\noindent \textbf{RESULTS} \\
\noindent \textbf{Photons in nonlinear cavity} \\
Figure~\ref{fig:1}A illustrates our experimental setup, comprising a Fabry-P\'{e}rot cavity filled with olive oil. Figure~\ref{fig:1}B shows white light transmission measurements through our cavity. The observed parabolic dispersion indicates that cavity photons behave as free particles with an effective mass $m^{\ast}=2.064 \times 10^{-35}$ kg. The ground state energy depends on the cavity length, which we control with a piezoelectric actuator holding one of our mirrors. Using a 532 nm continuous wave laser, we launch photon fluids with density and direction determined by the incident power and angle, respectively. The oil mediates effective photon-photon interactions in the  fluid. These interactions are repulsive because the refractive index of the oil decreases with increasing temperature and intensity, and they are non-instantaneous and non-local because of thermal relaxation and diffusion. Repulsive interactions modify the cavity transmission most significantly when the laser is blue-detuned from the ground state. In particular, we observe optical hysteresis and bistability (two steady states at a single driving condition) when scanning the laser power as shown in Fig.~\ref{fig:1}C. Bistability is closely related to superfluidity in optical cavities. At the minimum driving power needed for bistability, the excitation spectrum in the upper bistability branch is identical to the Bologiubov dispersion of an equilibrium superfluid~\cite{Carusotto04}. As the driving power increases, the excitation spectrum becomes increasingly gapped and superfluidity becomes more robust to fluctuations in light intensity. \\

\begin{figure*}[!t]
	\includegraphics[width=\textwidth]{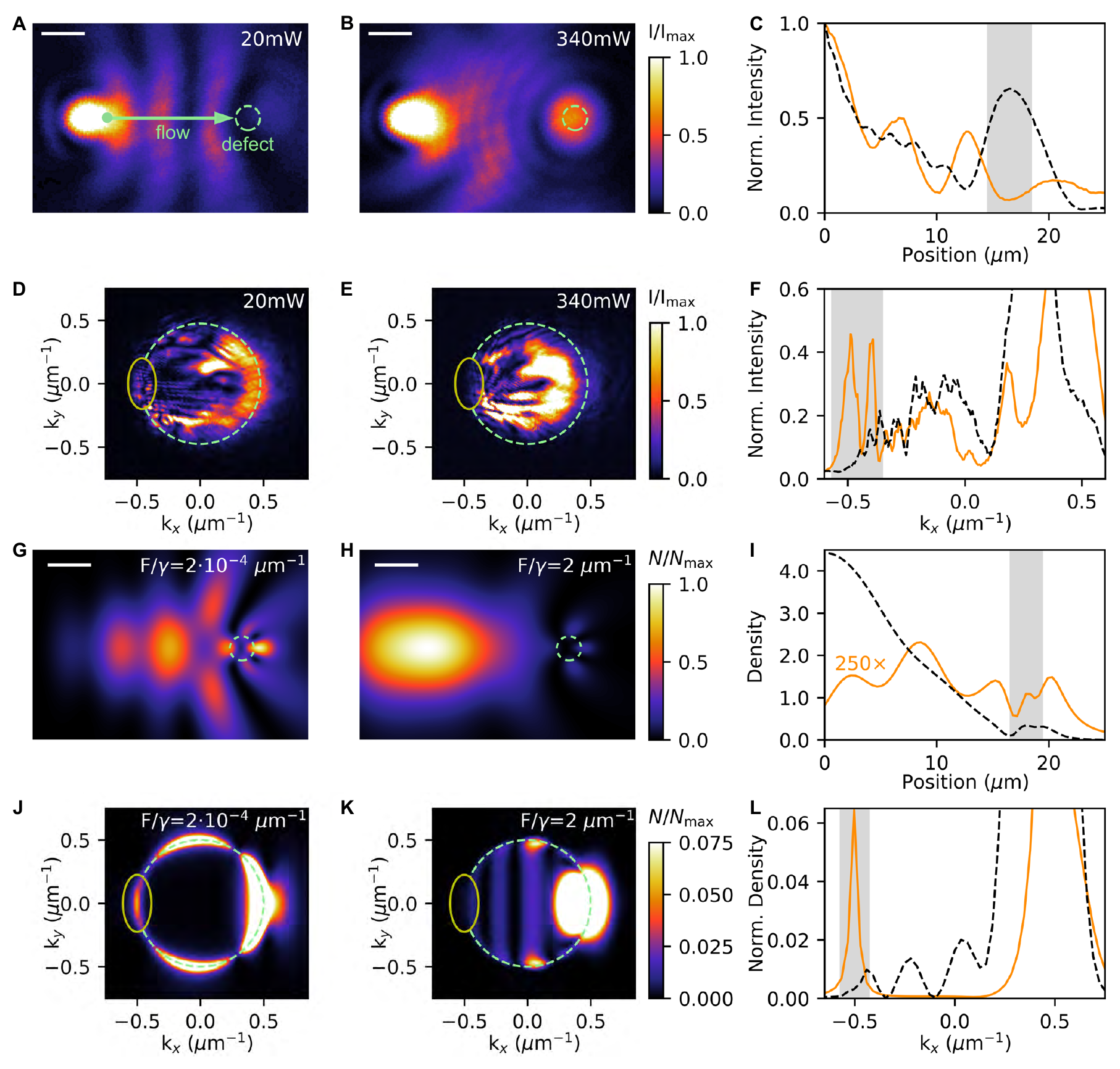}
	\caption{\label{fig:2}\textbf{Superfluid photons above a critical density.} Experimental transmitted intensity as a function of position in (\textbf{A}) and (\textbf{B}), and momentum in (\textbf{D}) and (\textbf{E}), for a laser power indicated in each panel. (\textbf{C}) Cuts along the flow axis passing through the center of the laser spot and the center of the defect in (A) and (B). The peaks observed at 20 mW but not at 340 mW are the result of scattering from the defect, which interferes with the incident flow and generates spatial fringes. (\textbf{F}) Cuts along $k_y=0$ in (D) and (E). The peaks observed at 20 mW inside the solid-line ellipse in (D), and in the shaded region in (F), evidence backscattering from the defect; these peaks disappear at 340 mW. (\textbf{G} to \textbf{L}) Numerical calculations as described in Methods, for two different laser amplitudes $F$ referenced to the cavity loss rate $\gamma$. $F/\gamma$ is such that fluid density is well-below and well-above the threshold for bistability in (G) and (J), and (H) and (K), respectively. Consequently, interference fringes are present in (G) but not in (H), and backscattering states inside the solid-line ellipse are present in (J) but not in (K). As for the experimental data, the cuts in (I) are taken along the flow axis passing through the center of the laser spot and the center of the defect in (G) and (H), and the cuts in (L) are taken along $k_y=0$ in (J) and (K). Scale bars indicate 5 $\mu$m. Solid dot in (A) corresponds to the position of 0 $\mu$m in (C). Dashed circles in (A, B) indicate approximate location of defect. Dashed circles in (G, H) indicate the location of the defect in simulations. The cuts in (C, I) are integrated vertically over the laser spot size and in (F, L) over $\Delta k_y=0.06$ $\mu$m$^{-1}$. In (C) data has been normalized to intensity at 0 $\mu$m. In (F, L) data for both powers has been normalized to the intensity at the peak of the incident momentum. Solid and dashed lines in (C, F, I, L) correspond to low and high power, respectively.}
\end{figure*}

\noindent \textbf{Suppression of backscattering from defect} \\
Figures~\ref{fig:2}A-F show measurements of a photon fluid propagating towards a naturally occurring defect in one of our mirrors. We consider two laser powers, well-below and well-above bistability. For low power, Fig.~\ref{fig:2}A shows fringes due to interference between the incident flow and scattering from the defect.   For high power (Fig.~\ref{fig:2}B), the fringes disappear and the defect becomes brighter. Figure~\ref{fig:2}C shows cuts of Figs.~\ref{fig:2}A,B along the flow direction, integrated over a vertical range matching the laser spot size. The interference fringes disappear at high power because of suppressed backscattering from the defect,  as confirmed by momentum space measurements in Figs.~\ref{fig:2}D-F.  Figures~\ref{fig:2}D,E show how the Rayleigh scattering ring (dashed circle) partially collapses, and backscattering is suppressed (see inside the solid-line ellipse), for increased power. Figure~\ref{fig:2}F shows cuts of Figs.~\ref{fig:2}D,E along $k_y=0$, integrated over $\Delta k_y=0.06$ $\mu$m$^{-1}$.  The shaded region in Fig.~\ref{fig:2}F highlights the backscattering states which disappear for increased power.

The partial collapse of the Rayleigh ring in Fig.~\ref{fig:2}E and the faint ripples in Fig.~\ref{fig:2}B indicate that scattering is not fully suppressed at high power. We attribute this to the broad momentum distribution of the incident laser and the finite momentum linewidth of cavity photons. The incident momentum distribution is shaped by the numerical aperture of the excitation objective, and by the slightly off-axis illumination (see Fig.~\ref{fig:1}A). This excitation scheme generates flow in every direction, but only states in and around the peak of the momentum distribution ($k_x=0.4 \mu$m$^{-1}$, $k_y=0$) flow straight into the defect. Since these are the most populated states, they establish an interaction energy that is greater along their trajectory than elsewhere in the cavity. Consequently, these more-populated states reach superfluidity at a laser power for which other states do not.

A distinctive feature of our measurements is the brightening of the defect observed at high power in Fig.~\ref{fig:2}B. In contrast, previous experiments with polariton fluids displayed a dark defect at all powers. This difference is due to the attractive nature of our defect, which corresponds to a potential well for the photon fluid. At low powers, scattering by the defect is proportional to the depth of the well. At high powers, scattering is suppressed and superfluid photons flow into the well.  The depth of the well and the short photon lifetime prevent the superfluid from escaping the well and propagating further. This results in a bright defect. We confirm this interpretation via numerical calculations based on a driven-dissipative Gross-Pitaevskii equation for photons coupled to a thermal field, presented next.

Figures~\ref{fig:2}G-L show numerical results for the photon density $N$, obtained using parameter values corresponding to our experiments (see Methods). We assumed the defect to be a Gaussian-shaped well of width 3.5 $\mu$m.  The simulations qualitatively reproduce our experimental observations. The suppression of interference fringes and the brightening of the defect  at high power are shown in Figs.~\ref{fig:2}G-I.  The partial collapse of the Rayleigh ring and the suppression of backscattering at high power are shown in Figs.~\ref{fig:2}J-L. Besides these qualitative agreements, we also observe differences between experiments and simulations. Our main source of  uncertainty is the shape and depth of the experimental defect, which strongly influence the fluid density $N$ near the defect. Another source of uncertainty concerns the modification of the cavity output rate at the defect.  Recall that the experimental transmitted intensity is always proportional to the intra-cavity photon density times the output rate squared. Meanwhile, $N$ in the simulations is just the intracavity intensity. Since the output rate at the defect should be greater than in the planar part of the cavity, the defect in experiments  may appear brighter (relative to the incident flow) than it actually is. This can explain why the simulation shows a modest peak in $N$ at the location of the defect   (see the dashed black line in the shaded region in  Fig.~\ref{fig:2}I) while the peak in the experiments is greater (see the dashed black line in the shaded region in  Fig.~\ref{fig:2}C). In any case, we are certain that our defect is attractive based on its brightening. For repulsive defects the fluid density always displays a minimum at the defect. \\

\begin{figure}[!t]
	\includegraphics[width=\columnwidth]{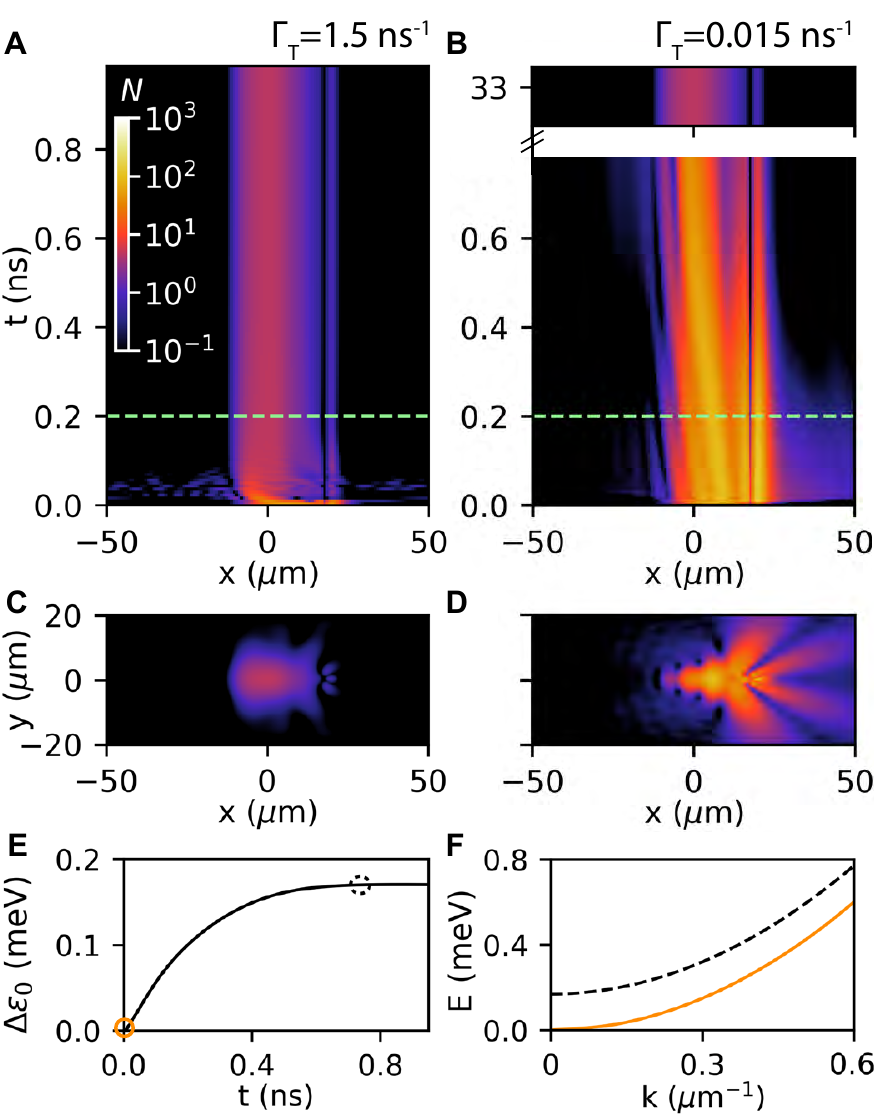}
	\caption{\label{fig:3}\textbf{From transient scattering to steady-state superfluidity.} (\textbf{A} and \textbf{B}) Time-dependent photon density $N$ along the flow axis passing through the center of the laser spot and the center of the defect. The thermal relaxation rate $\Gamma_T$ is $1.5$ ns$^{-1}$ in (A) and $0.015$ ns$^{-1}$ in (B). Consequently, a steady state is reached much faster in (A) than in (B). (\textbf{C} and \textbf{D}) Two-dimensional plots of $N$ at $t=0.2$ ns, indicated by the dashed lines in (A) and (B). After this time, the system has reached steady state for $\Gamma_T=1.5$ ns$^{-1}$ but not for $\Gamma_T=0.015$ ns$^{-1}$. Consequently, (C) displays the absence of interference fringes characteristic of superfluidity, while in (D) interference fringes are visible. (\textbf{E}) Time-dependent blueshift of excitation spectrum for $\Gamma_T=0.015$ ns$^{-1}$. (\textbf{F}) Spectrum of excitations at times indicated by corresponding circles in (E).}
\end{figure}

\noindent \textbf{Dynamics of the thermo-optical nonlinearity} \\
Our observation of superfluidity may seem surprising in view of the thermal relaxation time $\tau=16$ $\mu$s (see Ref.~\onlinecite{Geng_PRL2020}) being much longer than  the photon lifetime $\tau_{\mathrm{phot}}=4$ ps in our cavity. $\tau$ is the characteristic time in which the nonlinearity, and hence bistability, emerges in our system~\cite{Geng_PRL2020}. One may therefore expect the propagation of short-lived cavity photons to be unaffected by a slowly-rising nonlinearity. We tested this idea by calculating $N(t)$ in cavities with different thermal decay rates $\Gamma_T$, a parameter in our model which is inversely proportional to the experimental $\tau$ (see Methods). Results for two $\Gamma_T$ are shown in Fig.~\ref{fig:3}. In Figs.~\ref{fig:3}A,B we plot $N(t)$ along the horizontal axis passing through the center of the defect. For both $\Gamma_T$  we observe interference fringes due to scattering from the defect at short times. The duration of this transient scattering regime is commensurate with $\Gamma_T^{-1}$. After this transient regime, interference fringes disappear and steady-state superfluidity emerges.  In Figs.~\ref{fig:3}C,D we show the two-dimensional profile of $N$ at $t=0.2$ ns, indicated by dashed lines in Figs.~\ref{fig:3}A,B.  After $t=0.2$ ns, a steady state has been reached for $\Gamma_T=1.5$ ns$^{-1}$  in Figs.~\ref{fig:3}A,C, but not for $\Gamma_T=0.015$ ns$^{-1}$ in Figs.~\ref{fig:3}B,D. Consequently, Fig.~\ref{fig:3}D shows interference fringes due to scattering from the defect, while Fig.~\ref{fig:3}C does not. Clearly, while $\Gamma_T$ determines the duration of the transient scattering regime, the steady-state density profile reached after thermal relaxation is independent of $\Gamma_T$ in the range under consideration. For $\Gamma_T=0.015$ ns$^{-1}$, Fig.~\ref{fig:3}E shows the time-dependent blueshift of the excitation spectrum. The excitation spectrum at two instances is shown in Fig.~\ref{fig:3}F, clearly illustrating the formation of an energy gap. Such an energy gap in the excitation spectrum leads to superfluidity  as first realized by Landau~\cite{Landau1941}.\\

\begin{figure*}[!t]
	\includegraphics[width=\textwidth]{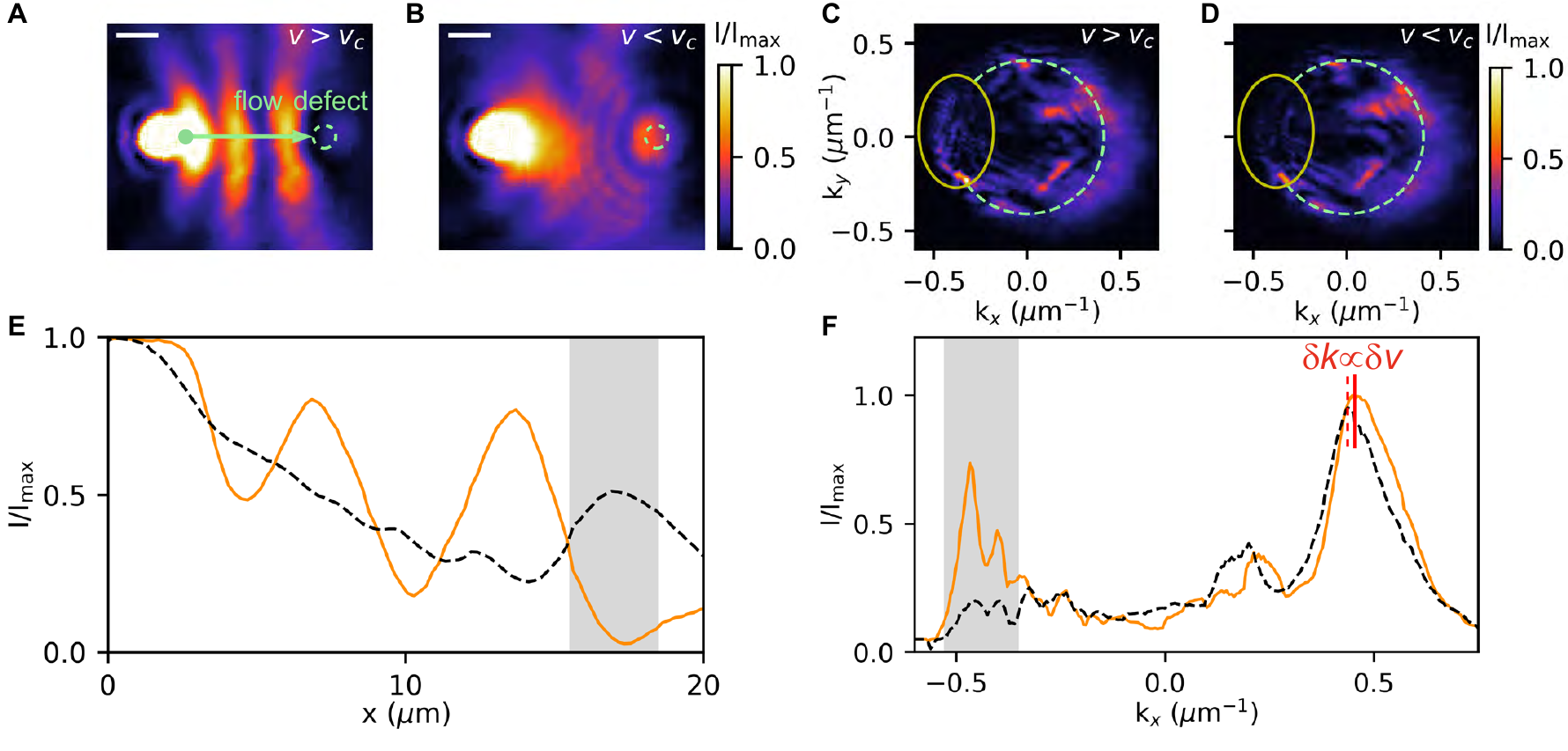}
	\caption{\label{fig:4}\textbf{Superfluidity of photons below a critical velocity.}  Experimental transmitted intensity as a function of position in (\textbf{A} and \textbf{B}), and momentum in (\textbf{C} and \textbf{D}), for slightly different cavity lengths corresponding to a 4\% change in the magnitude of the flow velocity $|v|$. The laser power is 360 mW for all measurements. Scale bars in (A) and (B) correspond to 5 $\mu$m. (\textbf{E}) Cuts along the flow axis passing through the center of the laser spot and the center of the defect in (A) and (B) as solid and dashed lines, respectively. The peaks observed for the larger $|v|$ (solid line) are due to scattering from the defect, which interferes with the incident flow and modulates the density. (\textbf{F}) Cuts along $k_y=0$ in (C) and (D) as solid and dashed lines, respectively. The peaks observed for the larger $|v|$ (solid line) inside the shaded region, which correspond to the peaks observed inside the solid-line ellipse in (C), evidence backscattering from the defect; these peaks disappear for the smaller $|v|$. Red vertical lines in (F) indicate the main incident wavevector, which changes only slightly in the viscous-to-superfluid transition. Solid dot in (A) corresponds to $x=0$ $\mu$m in (C). Dashed circles in (A, B) indicate approximate location of defect. The cuts in (E) are integrated vertically over the width of the laser spot and in (F) over $\Delta k_y=0.17$ $\mu$m$^{-1}$. In (E) the data for both velocities has been normalized to the intensity at $x=0$ $\mu$m, and in (F) to the intensity at the peak of the incident momentum of the solid-line cut.}
\end{figure*}

\noindent \textbf{Critical velocity} \\
Next we show the existence of a critical velocity above which superfluidity breaks down. The photon velocity is $v= \hbar k_{\parallel} /m^{\ast}$, with  $\hbar$ the reduced Planck constant and $k_{\parallel}$ the in-plane wavevector. Experimentally, we direct most of the light towards the defect by  illuminating the objective off-axis  (see Fig.~\ref{fig:1}A), such that the focused laser spot contains more wavevectors pointing at the defect than anywhere else. Meanwhile, we control the speed $|v|$ by adjusting the cavity length. This changes the radius $|k_{\parallel}|$ of the ring defined by the intersection of the photonic dispersion and the laser frequency. In this way, we probe scattering from the defect at different $|v| \propto |k_{\parallel}|$ and constant illumination conditions.

Fig.~\ref{fig:4} shows results for two speeds $|v|$  significantly larger than in Fig.~\ref{fig:2}. $|v|$ is slightly greater in Figs.~\ref{fig:4}A,C than in  Figs.~\ref{fig:4}B,D,   because the cavity is slightly longer in Figs.~\ref{fig:4}A,C. The change in cavity length corresponds to a change in energy detuning between the laser and the photonic ground state of only $\sim 40$ $\mu$eV, but this is sufficient to drastically alter the behavior. Notice the interference fringes due to scattering from the defect in Fig.~\ref{fig:4}A, which disappear in Fig.~\ref{fig:4}B.
Without scattering, light flows into the potential well in Fig.~\ref{fig:4}B. Then, the short cavity photon lifetime and the depth of the well prevents light from escaping the defect and propagating further. Consequently, the defect brightens.
Momentum-space measurements in Fig.~\ref{fig:4}C display backscattering states inside the solid-line ellipse, which are absent in Fig.~\ref{fig:4}D.  To illustrate the suppression of interference fringes and backscattering  more clearly, in Figs.~\ref{fig:4}E,F we present cuts of the measurements in Figs.~\ref{fig:4}A-D. The cuts are made along the flow axis. The results demonstrate that, in between the two velocities considered in Fig.~\ref{fig:4}, there exists a critical velocity $v_c$ for superfluidity. The two velocities considered in Fig.~\ref{fig:4} are in fact only slightly different. This is evident in the approximately equal radius of the Rayleigh ring in Figs.~\ref{fig:4}C,D, and  in the cuts shown in Fig.~\ref{fig:4}F. Notice how a tiny shift in the peak of the momentum distribution of the forward-propagating photon fluid (labeled as $\delta k$ in  Fig.~\ref{fig:4}F) brings about a drastic change in the back scattering, corresponding to the transition from viscous to superfluid photon flow. The abruptness of the transition is further illustrated in supplementary movie S1.

Finally, we note that varying the laser-cavity detuning changes the coupling of the excitation to the cavity mode. If the incident intensity on the defect drops below the critical intensity due to this detuning change, a superfluid-to-viscous transition  could be observed. However, this interpretation is ruled out by the fact that the intensity incident on the defect is actually greater when the incident wavevector is greater. This can be recognized by comparing the two peaks around $k_x=0.45$ $\mu$m$^{-1}$ in Fig.~\ref{fig:4}F. Hence, the superfluid-viscous transition observed in Fig.~\ref{fig:4} is indeed associated with the existence of a critical velocity and not with a change in fluid density. \\

\noindent \textbf{DISCUSSION AND CONCLUSION} \\
In summary, we have demonstrated signatures of superfluid cavity photons in  steady state  and at room temperature. Effective photon-photon interactions necessary for superfluidity are mediated by the thermo-optical nonlinearity of our oil-filled cavity. While thermal dynamics determine the time needed to reach steady state, a superfluid density profile is thereafter observed. A slow  thermo-optical nonlinearity, usually regarded as detrimental, opens up opportunities for probing photon fluids in new regimes and discovering new physics. For instance, a sudden removal of the driving laser in the superfluid regime is expected to generate a transient wake of moving optical vortices~\cite{Bolda01}. This prediction --- one of the first in the field --- has never been experimentally confirmed. This is likely due to the ultrafast nature of this effect in a cavity with instantaneous nonlinearity (e.g. a polariton system). In contrast, in our thermo-optical cavity with slow nonlinearity, the dynamics of the superfluid-to-viscous transition could be imaged using a standard fast camera. The slow superfluid dynamics in our system, in combination with the possibility of dynamically controlling the input laser amplitude and noise, opens up many opportunities for new research. For example, by ramping the laser amplitude at a rate comparable to the thermal relaxation rate, we may discover new universal scaling behavior as recently found in a single-mode thermo-optical nonlinear cavity~\cite{Geng_PRL2020}. Furthermore, by adding controlled amounts of noise to the input laser, one may probe the effects of fluctuations on photon fluid transport. Such effects may include non-Markovian stochastic-resonance-like phenomena~\cite{Peters20}, and  noise-assisted transport~\cite{Plenio08}.   Finally, beyond the physics of a single defect in a  planar potential landscape,  our system can be readily extended to probe the physics of photon fluids in complex potential landscapes.    This could be done by micro-structuring one of our cavity mirrors using focused-ion beam milling~\cite{Trichet15} or direct laser writing ~\cite{Weitz19, Kurtscheid_2020}.\\

\noindent \textbf{METHODS} \\
\noindent \textbf{Experimental design} \\
Our Fabry-P\'{e}rot cavity is made by two distributed Bragg reflectors (DBRs) with a peak reflectance of $99.9$\% at $530$ nm. One of the mirrors is fixed. The other mirror is mounted on a six degree-of-freedom positioner used to control its position and orientation with a precision of 1 nm and 1 micro-degree, respectively. All experimental results except those in Fig.~\ref{fig:1}b were obtained with a single-mode, continuous wave (CW), laser emitting at 532 nm. Both our excitation and collection objectives have a magnification of 20$\times$ and a numerical aperture of 0.4. The transmitted light is either measured as a function of time using a photodetector and an oscilloscope, or as a function of position and momentum using cameras. In all of our measurements we probed the $6^{\textrm{th}}$ longitudinal cavity mode. For the measurements in Fig.~\ref{fig:1}c, the laser power is modulated at 40 Hz.  \\

\noindent \textbf{Numerical calculations} \\
For the theoretical description of the oil-filled cavity, we used coupled equations for the light field and the spatial temperature distribution. The equation of motion for the photon field $\psi$ reads (in units of $\hbar=1$)
\begin{equation}
\begin{split}
i \dot{ \psi}(\mathbf{r},t) = &\left (\epsilon_0-\frac{i}{2} \gamma-\frac{\nabla^2}{2m^{\ast}}
+ V_D(\mathbf{r}) + \alpha \Delta T(\mathbf{r},t) \right) \\
& \times  \psi(\mathbf{r},t) + F_L(\mathbf{r}) e^{-i \omega_L t+i \mathbf{k}_L \cdot \mathbf{r}}.
\label{eq:gpe}
\end{split}
\end{equation}
Here $\epsilon_0$ is the cavity resonance frequency at room temperature, $m^*$ is the effective photon mass,  $\gamma$ the cavity line width and $V_D$ is the defect potential due to the local deformation of one of the mirrors. The deviation of the temperature from room temperature $\Delta T$ leads to a change in the cavity frequency through the thermo-optic coefficient $\alpha$. The coherent laser excitation has amplitude $F_L$, frequency $\omega_L$ and wavevector $\mathbf{k}_L$.
The temperature dynamics satisfy
\begin{equation}
\begin{split}
\dot{ \Delta T} (\mathbf{r},t) =& - \Gamma_T  \Delta T(\mathbf{r},t) + D \nabla^2  \Delta T(\mathbf{r},t)\\ & + \beta |\psi(\mathbf{r},t)|^2.
\label{eq:T}
\end{split}
\end{equation}
Here $\Gamma_T$ is the relaxation rate of the temperature in the transverse direction due to thermal conduction of heat through the mirrors. $D$ is the in-plane thermal diffusion constant and $\beta$ describes the heating due to photon absorption. The associated photon number decay is included in the photon loss rate $\gamma$. In order to model the experiments, we have numerically solved these coupled equations with a split step method.

The effective photon-photon interaction constant for the bistability in a homogeneous system can be derived by setting the time derivative of the temperature equal to zero, which yields $\Delta T = (\beta/\Gamma_T) |\psi|^2$. Plugging this result back in Eq. \eqref{eq:gpe}, one sees that the effective photon nonlinearity is $g_{\rm eff} =  (\alpha \beta/\Gamma_T)$.\\

\vspace{5mm}

\noindent \textbf{SUPPLEMENTARY MATERIAL} \\
\noindent  Movie S1. Superfluidity of photons below a critical velocity. \\
\noindent Experimental transmitted intensity as a function of position (left panel) and momentum (right panel). As time progresses in the movie, the cavity length increases. Since the excitation laser has a fixed energy and a broad momentum distribution, an increasing cavity length results in an increasing flow velocity for cavity photons. The increasing flow velocity is clearly evidenced in the right panel, where the radius of the Rayleigh ring expands in time. Recall that the Rayleigh ring radius is proportional to the magnitude of the wavevector, and hence to the velocity, of cavity photons. The movie shows that, at low velocity, the photon fluid is in the superfluid regime. This is evidenced by the absence of interference fringes due to scattering from the defect (left panel) and the absence of backscattering (right panel, inside solid ellipse). Then, as the flow velocity increases beyond a critical velocity, a transition to the viscous regime occurs. This is evidenced by the emergence of fringes in real space (left panel) and of intensity peaks with momentum corresponding to backscattering by the defect (right panel, inside solid ellipse). Yellow ‘S’ and red ‘V’ indicate superfluid and viscous regime, respectively. Peak horizontal incident wavevector at $k_y=0$ $\mu$m$^{-1}$ is displayed above right panel. Scale bar represents 5 $\mu$m.


%

\vspace{5mm}

\noindent \textbf{Acknowledgments} \normalsize \\
\noindent This work is part of the research programme of the Netherlands Organisation for Scientific Research (NWO). We thank Ilan Shlesinger, Peter Kirton, and Alberto Amo for stimulating discussions, and Natalia Berloff, Pavlos Lagoudakis, and Pavlos Savvidis for organizing the 2019 HPM conference which stimulated our collaboration and led to the present results.  S.R.K.R. acknowledges an ERC Starting Grant with project number 852694 and a NWO Veni grant with file number 016.Veni.189.039. \\

\noindent {\textbf{Author contributions} \normalsize \\
\noindent G.K. performed the experiments, with critical contributions from  S.R.K.R., Z.G., and K.J.H.P.. M. W. performed the simulations. S.R.K.R. conceived the experiments and supervised the work. S.R.K.R. and G.K wrote the manuscript with critical contributions from M.W.. All authors discussed the results and the manuscript. \\

\end{document}